\documentclass[aps,prd,amsmath,amssymb,
twocolumn,showpacs,superscriptaddress,floatfix]{revtex4-1}

\usepackage{graphicx}% Include figure files
\usepackage{dcolumn}% Align table columns on decimal point
\usepackage{bm}% bold math
\usepackage{hyperref}% add hypertext capabilities
\usepackage{verbatim}% comment
\usepackage[mathlines]{lineno}% Enable numbering of text and display math
\usepackage{array}
\newcolumntype{P}[1]{>{\centering\arraybackslash}p{#1}} % center table cells horizontally 
\newcolumntype{M}[1]{>{\centering\arraybackslash}m{#1}} % Center table cells vertically and horizontally

\begin{document}

\title{Underlying event measurements in $p$+$p$ collisions at $\sqrt{s}= 200 $ GeV at RHIC}% Force line breaks with \\

\affiliation{Abilene Christian University, Abilene, Texas   79699}
\affiliation{AGH University of Science and Technology, FPACS, Cracow 30-059, Poland}
\affiliation{Alikhanov Institute for Theoretical and Experimental Physics NRC "Kurchatov Institute", Moscow 117218, Russia}
\affiliation{Argonne National Laboratory, Argonne, Illinois 60439}
\affiliation{American University of Cairo, New Cairo 11835, New Cairo, Egypt}
\affiliation{Brookhaven National Laboratory, Upton, New York 11973}
\affiliation{University of California, Berkeley, California 94720}
\affiliation{University of California, Davis, California 95616}
\affiliation{University of California, Los Angeles, California 90095}
\affiliation{University of California, Riverside, California 92521}
\affiliation{Central China Normal University, Wuhan, Hubei 430079 }
\affiliation{University of Illinois at Chicago, Chicago, Illinois 60607}
\affiliation{Creighton University, Omaha, Nebraska 68178}
\affiliation{Czech Technical University in Prague, FNSPE, Prague 115 19, Czech Republic}
\affiliation{Technische Universit\"at Darmstadt, Darmstadt 64289, Germany}
\affiliation{ELTE E\"otv\"os Lor\'and University, Budapest, Hungary H-1117}
\affiliation{Frankfurt Institute for Advanced Studies FIAS, Frankfurt 60438, Germany}
\affiliation{Fudan University, Shanghai, 200433 }
\affiliation{University of Heidelberg, Heidelberg 69120, Germany }
\affiliation{University of Houston, Houston, Texas 77204}
\affiliation{Huzhou University, Huzhou, Zhejiang  313000}
\affiliation{Indian Institute of Science Education and Research (IISER), Berhampur 760010 , India}
\affiliation{Indian Institute of Science Education and Research (IISER) Tirupati, Tirupati 517507, India}
\affiliation{Indian Institute Technology, Patna, Bihar 801106, India}
\affiliation{Indiana University, Bloomington, Indiana 47408}
\affiliation{Institute of Modern Physics, Chinese Academy of Sciences, Lanzhou, Gansu 730000 }
\affiliation{University of Jammu, Jammu 180001, India}
\affiliation{Joint Institute for Nuclear Research, Dubna 141 980, Russia}
\affiliation{Kent State University, Kent, Ohio 44242}
\affiliation{University of Kentucky, Lexington, Kentucky 40506-0055}
\affiliation{Lawrence Berkeley National Laboratory, Berkeley, California 94720}
\affiliation{Lehigh University, Bethlehem, Pennsylvania 18015}
\affiliation{Max-Planck-Institut f\"ur Physik, Munich 80805, Germany}
\affiliation{Michigan State University, East Lansing, Michigan 48824}
\affiliation{National Research Nuclear University MEPhI, Moscow 115409, Russia}
\affiliation{National Institute of Science Education and Research, HBNI, Jatni 752050, India}
\affiliation{National Cheng Kung University, Tainan 70101 }
\affiliation{Nuclear Physics Institute of the CAS, Rez 250 68, Czech Republic}
\affiliation{Ohio State University, Columbus, Ohio 43210}
\affiliation{Panjab University, Chandigarh 160014, India}
\affiliation{Pennsylvania State University, University Park, Pennsylvania 16802}
\affiliation{NRC "Kurchatov Institute", Institute of High Energy Physics, Protvino 142281, Russia}
\affiliation{Purdue University, West Lafayette, Indiana 47907}
\affiliation{Rice University, Houston, Texas 77251}
\affiliation{Rutgers University, Piscataway, New Jersey 08854}
\affiliation{Universidade de S\~ao Paulo, S\~ao Paulo, Brazil 05314-970}
\affiliation{University of Science and Technology of China, Hefei, Anhui 230026}
\affiliation{Shandong University, Qingdao, Shandong 266237}
\affiliation{Shanghai Institute of Applied Physics, Chinese Academy of Sciences, Shanghai 201800}
\affiliation{Southern Connecticut State University, New Haven, Connecticut 06515}
\affiliation{State University of New York, Stony Brook, New York 11794}
\affiliation{Temple University, Philadelphia, Pennsylvania 19122}
\affiliation{Texas A\&M University, College Station, Texas 77843}
\affiliation{University of Texas, Austin, Texas 78712}
\affiliation{Tsinghua University, Beijing 100084}
\affiliation{University of Tsukuba, Tsukuba, Ibaraki 305-8571, Japan}
\affiliation{United States Naval Academy, Annapolis, Maryland 21402}
\affiliation{Valparaiso University, Valparaiso, Indiana 46383}
\affiliation{Variable Energy Cyclotron Centre, Kolkata 700064, India}
\affiliation{Warsaw University of Technology, Warsaw 00-661, Poland}
\affiliation{Wayne State University, Detroit, Michigan 48201}
\affiliation{Yale University, New Haven, Connecticut 06520}

\author{J.~Adam}\affiliation{Brookhaven National Laboratory, Upton, New York 11973}
\author{L.~Adamczyk}\affiliation{AGH University of Science and Technology, FPACS, Cracow 30-059, Poland}
\author{J.~R.~Adams}\affiliation{Ohio State University, Columbus, Ohio 43210}
\author{J.~K.~Adkins}\affiliation{University of Kentucky, Lexington, Kentucky 40506-0055}
\author{G.~Agakishiev}\affiliation{Joint Institute for Nuclear Research, Dubna 141 980, Russia}
\author{M.~M.~Aggarwal}\affiliation{Panjab University, Chandigarh 160014, India}
\author{Z.~Ahammed}\affiliation{Variable Energy Cyclotron Centre, Kolkata 700064, India}
\author{I.~Alekseev}\affiliation{Alikhanov Institute for Theoretical and Experimental Physics NRC "Kurchatov Institute", Moscow 117218, Russia}\affiliation{National Research Nuclear University MEPhI, Moscow 115409, Russia}
\author{D.~M.~Anderson}\affiliation{Texas A\&M University, College Station, Texas 77843}
\author{A.~Aparin}\affiliation{Joint Institute for Nuclear Research, Dubna 141 980, Russia}
% ...

\collaboration{STAR Collaboration}\noaffiliation

\date{\today}% It is always \today, today,

\begin{abstract}
Particle production sensitive to non-factorizable and non-perturbative processes that contribute to the underlying event associated with a high transverse momentum ($p_{T}$) jet in proton+proton collisions at $\sqrt{s}$=200 GeV is studied with the STAR detector. Each event is divided into three regions based on the azimuthal angle with respect to the highest-$p_{T}$ jet direction:  in the leading jet direction (``Toward"), opposite to the leading jet (``Away"), and perpendicular to the leading jet (``Transverse"). In the Transverse region, the average charged particle density is found to be between 0.4 and 0.6 and the mean transverse momentum, $\langle  p_{T}\rangle$, between 0.5-0.7~GeV/$c$ for particles with $p_{T}$$>$0.2~GeV/$c$ at mid-pseudorapidity ($|\eta|$$<$1) and jet $p_{T}$$>$15~GeV/$c$. Both average particle density and $\langle p_{T}\rangle$ depend weakly on the leading jet $p_{T}$. 
Closer inspection of the Transverse region hints that contributions to the underlying event from  initial- and final-state radiation are significantly smaller in these collisions than at the higher energies, up to 13 TeV, recorded at the LHC. Underlying event measurements associated with a high-$p_{T}$ jet will contribute to our understanding of QCD processes at hard and soft scales at RHIC energies, as well as provide constraints to modeling of underlying event dynamics.

\end{abstract}

\maketitle

\section{\label{sec:intro}Introduction}
Understanding the underlying physics of hadronic collisions requires detailed characterization of the particle production processes. Proton+proton ($p$+$p$) collisions include elastic and inelastic scatterings, with inelastic $p$+$p$ scatterings consisting of single diffractive, double diffractive and non-diffractive processes. In non-diffractive events, when a hard scattering occurs with large momentum transfer ($p_{T}$$\geq$2~GeV/$c$) from the longitudinal to the transverse plane,  other processes could occur in addition to the production of a high energy dijet. These  additional processes
include softer secondary hard scatterings or multiple parton interactions (MPI), gluon radiation or quark-antiquark splittings from the initial- or final-state partons of the primary hard scattering (ISR/FSR), and color reconnections with the beam-remnant partons (BR).
Cumulatively, these result in what is referred to, in experiments,  as the underlying event.  The partons produced in these processes reduce their virtuality and finally fragment mainly into low-energy particles. The properties of the  dijet produced in the initial hard scattering can be calculated by perturbative QCD, with good agreement between experimental measurements and theoretical calculations after careful correction for the underlying event activity~\cite{CMS:JetSpectraPP276TeV,ATLAS:JetFragPP7TeV,Adam:2019aml}. Modeling the soft physics, which dominates the underlying event activity, is challenging since it does not factorize and requires non-perturbative calculations \cite{PYTHIA6tune_arxiv1005.3457, Skands:2010ak}. Experimental studies of the underlying event activity, spanning non-perturbative and perturbative QCD and including sensitivities to multi-scale physics, can help us to improve theoretical modeling and understand the QCD processes. 
Underlying event activity is often experimentally accessed through topological structure observables, such as particle production away from the primary hard scattering reference direction. The CDF collaboration used the highest-$p_{T}$ charged particle, leading jet, or Drell-Yan pair in each proton+antiproton event to define the hard scattering reference \cite{CDF_PRD92,CDF_arxiv1003.3146}. At LHC energies, the ALICE, ATLAS, and CMS collaborations have used the highest-$p_{T}$ charged particle, charged jet, jet with both charged and neutral particles, or $Z$ boson as the hard scattering references \cite{CMS_UE_arxiv1507.07229, ATLAS_UE_arxiv1701.05390, ATLAS_Z_UE_arXiv1409.3433, ALICE:2011ac, CMS_DY_UE_arxiv1204.1411}. In this analysis, we use the highest-$p_{T}$ jet as our hard scattering reference. 

At TeV collision energy scales, the average mid-pseudorapidity charged multiplicity density and mean transverse momentum, both  sensitive to the underlying event activity, were observed to be positively correlated with observables sensitive to the hard scattering energy~\cite{CDF_arxiv1003.3146,CMS_UE_arxiv1507.07229, ATLAS_UE_arxiv1701.05390, ATLAS_Z_UE_arXiv1409.3433, ALICE:2011ac, CMS_DY_UE_arxiv1204.1411}. These positive correlations are understood as resulting from increasing contributions from wide angle ISR/FSR, as the transferred momentum scale, Q$^2$, of the hard scattering increases. However, at forward rapidity, CMS found that the relationship between underlying event particle production and leading jet $p_{T}$ depends on collision energy \cite{CMS_fowardUE_JHEP04_2013_072}: at $\sqrt{s}$ = 0.9 TeV, the underlying event  particle production activity and leading jet $p_{T}$ were negatively correlated, while at 7 TeV they were mostly positively correlated for leading charged jets with 1$<$$p_{T}$$<$15~GeV/$c$. The negative correlations reported at the lower collision energy could be due to energy conservation constraining the underlying event production. For these reasons, it is worth exploring the underlying event at RHIC with even lower energies to search for its relationship with the leading jet energy at mid-pseudorapidity.

\section{\label{sec:exp}Experiment and Data Analysis}

In this paper, we report  measurements of underlying event activity in $p$+$p$ collisions at $\sqrt{s} $ = 200~GeV by the STAR experiment at RHIC. The data used in this analysis were collected in 2012 with an integrated luminosity of $\sim$23~pb$^{-1}$. The major subsystems used for the analysis were the Time Projection Chamber (TPC) \cite{TPC}, and the Barrel Electromagnetic Calorimeter (BEMC) \cite{BEMC}. 
The TPC provides charged particle tracking with good momentum resolution, while the 4800 isolated towers of the BEMC record the energy deposited by photons, electrons, $\pi^{0}$, and $\eta$ mesons. Both cover mid-pseudorapidity ($|\eta|$$<$1) with full azimuthal angle ($\phi$) coverage.

This analysis used both a minimum bias and a calorimeter-triggered dataset. The minimum bias data required a coincidence of signals from  the Vertex Position Detectors \cite{VPD}, which measure photons from $\pi^{0}$ decays at forward and backward pseudorapidity, 4.2$\leq$$|\eta|$$\leq$5.1. The calorimeter jet-patch triggered data required a minimum transverse energy ($E_{T}$) in a $\Delta\eta\times\Delta\phi$ region of $\approx 1 \times 1$ of the BEMC.  In this analysis, three $E_{T}$ trigger thresholds were used: 3.5, 5.4, and 7.3~GeV. Jets reconstructed from the calorimeter-triggered dataset are therefore biased towards  higher neutral energy fractions \cite{STARJNF_arxiv1405.5134}. Corrections for this electromagnetic calorimeter-trigger bias are determined via a data-driven technique using the minimum bias data, and only the corrected calorimeter-triggered data are reported in the final results.  
Event-level and particle-level selections were applied to the recorded data. The events were required to have a reconstructed primary event vertex, $z_{vtx}$, within 30 cm of the center of the TPC along the beam axis ($z$) in order to ensure nearly uniform detector acceptance. Events containing a charged particle with $p_{T}$$>$20~GeV/$c$ were discarded to avoid events with low tracking momentum reconstruction resolution. Events with deposited energy in a single BEMC tower of $E_{T}$$>$20~GeV  were also discarded for symmetry. The charged particle tracks reconstructed in the TPC were required to satisfy the following conditions: 0.2$<$$p_{T}$$<$20~GeV/$c$ for high tracking efficiency; a distance of closest approach (dca) to the event vertex of $|dca|$$< $1 cm to ensure particles are from the primary collision vertex; a number of fit points along the track greater than 20 out of a maximum of 45; and a ratio of the number of fit points  to the maximum number of possible fit points larger than 0.52 for good primary track reconstruction \cite{TPC,CUTSEXPLAINED}. The neutral energy deposits in each BEMC tower were required to have 0.2$<$$E_{T}$$<$20~GeV. For the particles used in this paper, the pseudorapidity region was restricted to $|\eta|$$<$1.

The simulation sample used for detector response and background correction in this analysis was generated using PYTHIA 6.4.28 \cite{PYTHIA6_Sjostrand:2006za} with the CTEQ6L1 PDF \cite{PDF_Pumplin:2002vw} and the Perugia 2012 tune \cite{PYTHIA6tune_arxiv1005.3457} with the PARP(90) parameter changed to 0.213, 
which is, throughout this article, referred to as PYTHIA 6 (STAR) \cite{Adam:2019aml}.  PYTHIA 6 (STAR) is tuned with STAR published minimum bias identified particle spectra \cite{STAR_PLB616_8_16,STAR_PRL108_072302} resulting in agreement with the inclusive identified charged pion cross-sections at the 10\% level~\cite{Kevin_diss}. The PYTHIA parameter PARP(90) is related to the energy scaling of the minimum bias and underlying event phenomena \cite{PYTHIA6tune_arxiv1005.3457}. PYTHIA 6 (STAR) generated events are processed through the STAR GEANT3 \cite{GEANT3_Brun:1987ma} detector simulation with $\pi^{0}$, $\eta$, and $\Sigma^{0}$ decays and weak decays handled by GEANT and set as stable in PYTHIA
. The simulated GEANT output events were further embedded into zero bias (randomly triggered) experimental data. This embedding procedure simulates a similar background, such as pile-up, beam-gas interactions and cosmic rays, as exists for the experimental measurements \cite{STARJNF_arxiv1405.5134}. 
The default PYTHIA 6.4.28 Perugia 2012 tune with the CTEQ6L1 PDF, and the default PYTHIA 8.215 Monash 2013 tune \cite{pythia8tune_arxiv1404.5630} simulations were also used for comparison with corrected data, later referred to as PYTHIA 6 and PYTHIA 8, respectively.

All charged tracks with $p_{T}$$>$0.2~GeV/$c$ and all BEMC towers with deposited $E_{T}$$>$0.2~GeV, within $|\eta|$$<$1 were used to reconstruct jets. Jet reconstruction used the anti-$k_{\text T}$ algorithm~\cite{ANTIKT} with a resolution parameter $R_{\text anti-k_{T}}$=0.6 as implemented in the FastJet package~\cite{Cacciari2012}. The reconstructed jets were restricted to  $|\eta_{\text jet}|$$<$0.4 to ensure that all tracks/neutral energy deposits of the jets were inside our detector acceptance. The jet neutral energy fraction \cite{STARJNF_arxiv1405.5134} was required to be less than 90\% to minimize non-collision backgrounds such as beam-gas interactions and cosmic rays. Jet axes were further required to be within $\Delta R=\sqrt{\Delta \eta ^2 + \Delta \phi ^2 }$=0.6 distance from the jet-patch center for calorimeter-triggered data \cite{STARJNF_arxiv1405.5134}. The reported jet energies include contributions from the underlying event and are corrected for detector effects including pile-up via the unfolding procedure described below.  

The analysis followed the CDF topological structure method \cite{CDF_PRD92}. For each collision event, the leading, highest $p_T$, jet azimuthal angle ($\phi_{ \text jet}$) is defined as the reference angle. The reconstructed charged particles are then categorized into different regions by the relative difference of their azimuthal angle ($\phi_{i}$) to the jet's reference angle, $\Delta \phi \equiv \phi_{i}-\phi_{\text jet}$. 
The ``Toward" region contains particles with $|\Delta\phi|$$<$60$^{\circ}$, while the ``Away" region, which is not required to contain a reconstructed jet, is defined as those particles with $|\Delta \phi$-180$^{\circ}|$$<$60$^{\circ}$. Finally, the ``Transverse" region covers
60$^o$$\le$$|\Delta \phi |$$\le $$120^{\circ}$.  The activity in each region is integrated over  $|$$\eta$$|$$<$1, which is inside of the TPC acceptance. 
The underlying event activity is then accessed through the $\eta$ integrated Transverse region and reported for charged tracks only. The average charged particle multiplicity density, $ \langle \frac{dN_{ch}}{d\eta d\phi} \rangle$, and the mean transverse momentum, $\langle p_{T}\rangle$, are studied as a function of the leading jet $p_{T}$. 

Each event has two Transverse regions, 60$^{\circ}$$\le$$\Delta\phi$$\le$120$^{\circ}$ and -120$^{\circ}$$\le$$\Delta\phi$$\le$-60$^{\circ}$.  For each event, the Transverse region with larger charged multiplicity density was defined as ``TransMax", and the other as ``TransMin". This binning procedure will cause the results to differ, even if the TransMax and TransMin distributions originate from the same parent distribution, due to region-to-region  statistical fluctuations.   
Since the MPI and BR processes are potentially unrelated to the hard scattering jet angle, both processes make comparable contributions to both the TransMax and TransMin regions. However, ISR/FSR may produce a wide angle third jet, the fragmentation particles of which have a high probability of being recorded in only one of the Transverse regions. ISR/FSR can therefore also result in a significant difference in the two regions' multiplicity densities. Comparisons of TransMax and TransMin regions may therefore reveal details of the various processes contributing to the underlying event. The TransMin region is sensitive to particle production from BR and MPI, while the TransMax region also contains signal from ISR/FSR.

The measured data were corrected for trigger bias, detector inefficiency, cosmic-ray background, and pile-up effects. The trigger bias was corrected for by weighting calorimeter-triggered data to have the same jet neutral fraction distribution as minimum bias data. The trigger bias corrections were $p_{T}$ dependent and were less than 15\% for the multiplicity density results and less than 20\% for $\langle p_{T}\rangle$. 

The calorimeter-triggered data were then unfolded using a detailed simulation of the trigger to correct for inefficiencies as a function of jet $p_T$. Differences to the previously mentioned bias-corrected result were accounted for as part of the systematic uncertainty estimate.
The detector inefficiency, resolution effects, and background corrections were performed using 2-dimensional Bayesian unfolding  \cite{UNFOLD}. 
The response matrices were constructed from matched generator-level and detector-level observables \cite{STAR_Dijetsigma_PRD_2017}. The generator-level is from PYTHIA 6 (STAR) and the detector-level is from the embedded sample to include the backgrounds to the signal, such as the pile-up contribution.  
A jet pair was considered matched if the distance of a leading jet at the detector-level to a leading or sub-leading jet at the generator-level was less than the jet resolution parameter, $R_{\text anti-k_{T}}$=0.6. When there was a match, the leading jet $p_{T}$ at the generator-level was used to create the response matrices. For the track level observable $\langle p_{T}\rangle$, a matched track pair was also required. The track pair is considered matched when more than ten hits are matched \cite{CUTSEXPLAINED}. The embedding sample had the same track and jet quality requirements applied as the experimental data which included requiring the particle $p_{T}$$>$0.2~GeV/$c$, $|\eta|$$<$1 and leading jet $|\eta_{jet}|$$<$0.4. 

\begin{figure}[htb]
\centering
\includegraphics[width=0.4\textwidth]{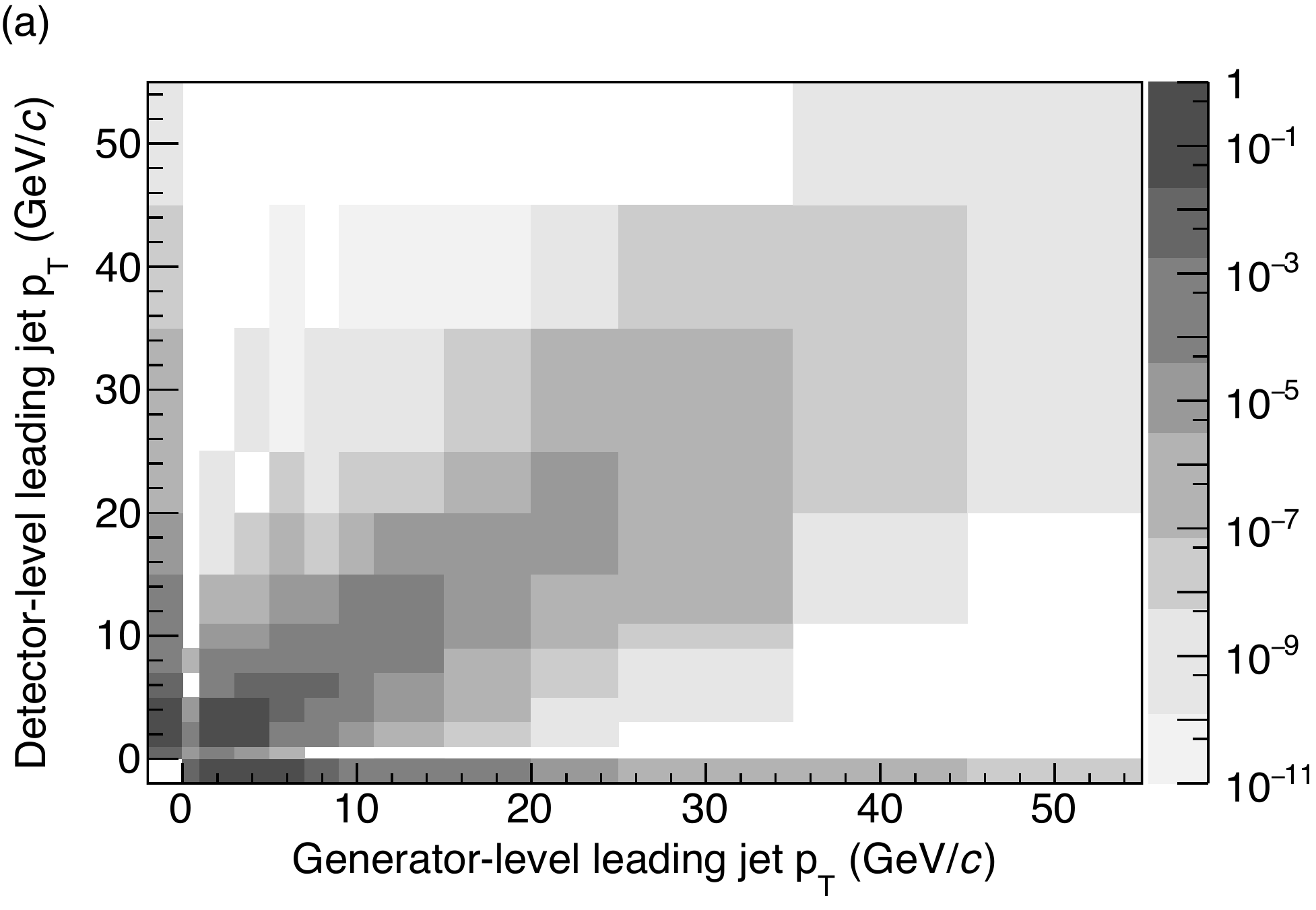}
\includegraphics[width=0.4\textwidth]{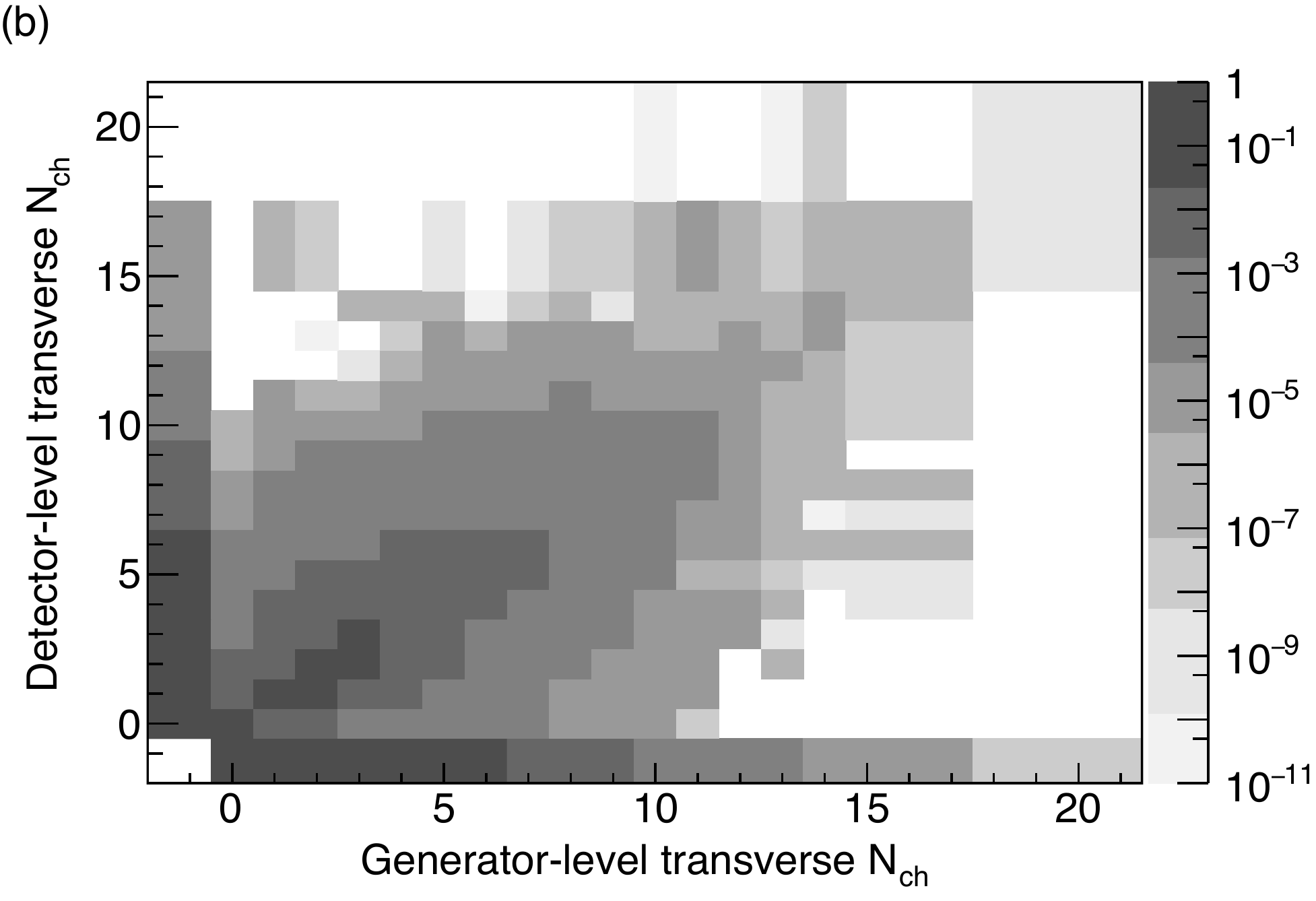}
\caption{Projections of the 4-dimensional response matrix for (a) the leading jet $p_{T}$ and (b) Transverse charged particle multiplicity $N_{ch}$, each with the other variable integrated. X-axis is the generator-level variable and the Y-axis is the detector-level variable. The bins along the negative axes are for the cases when a matched pair is not found. }
\label{fig:Unfold}
\end{figure}

Figure~\ref{fig:Unfold} shows 2-dimensional projections of the 4-dimensional response matrix. Panel (a) is the projection onto the leading jet $p_{T}$ and panel (b) is the Transverse charged multiplicity integrated over the leading jet $p_{T}$. The negative bins  represent the probability for no match to be identified for the concerned variable. Although the two figures exhibit a fairly linear correlation between the detector-level and generator-level variables, their correlations are smeared. To ensure closure of  the unfolding correction procedure, the simulation data were divided into two subsets, one was used for training and one for the unfolding closure test. This successful closure test also demonstrated that the uncertainty due to the response matrices is negligible. The unfolding method uncertainty was estimated by varying the Bayesian iterations (with 4 as the default), the unfolding prior, and the maximum tower energy cut of 20~GeV. The effects on the unfolded results due to the uncertainty on the TPC tracking efficiency  of 4\% and the BEMC tower energy calibration uncertainty of 4\% were also estimated. For both cases the response matrices were recalculated after varying the simulations by the appropriate uncertainty and the data unfolded with the new matrix. The systematic errors obtained from these variations were added quadratically with the unfolding uncertainty to form the total systematic uncertainty. The total correction uncertainties were found to be less than 15\%.

\section{\label{sec:result}Results and Discussion}

\begin{figure}[htb]
\centering
\includegraphics[width=0.4\textwidth]{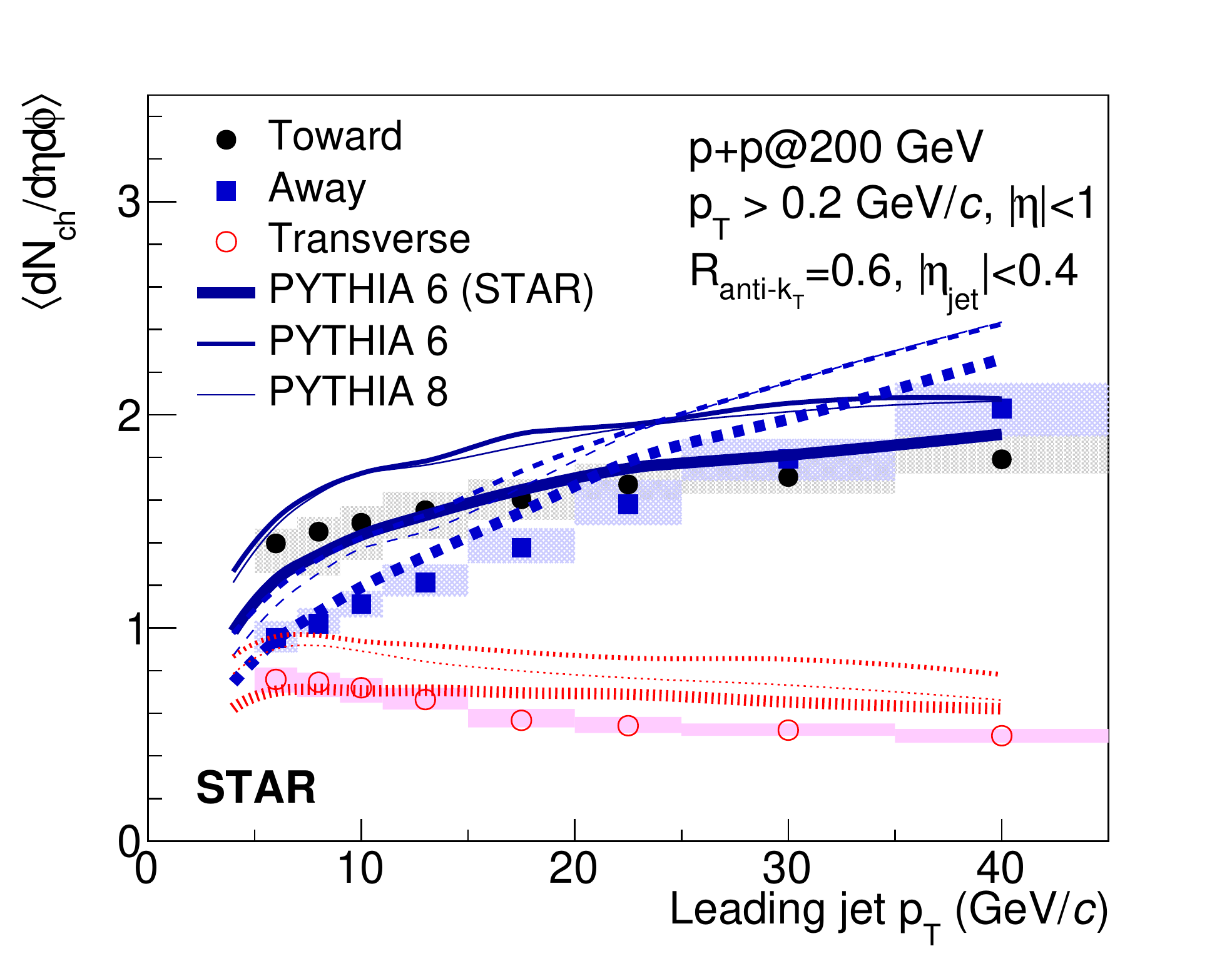}
\caption{Average charged particle multiplicity densities for Toward, Away, and Transverse regions as functions of the leading jet $p_{T}$, with charged particle $p_{T}$$>$0.2~GeV/$c$. The wide curves are PYTHIA 6 (STAR). The middle width curves are default PYTHIA 6 Perugia 2012 tune. The thin curves are PYTHIA 8 Monash 2013 tune.   The solid curves are the Toward region. The sparse dashed curves are the Away region. The dense dashed curves are the Transverse region. 
}
\label{fig:mult}
\end{figure}

Figure~\ref{fig:mult} shows the fully corrected average charged particle multiplicity densities for particles with $p_{T}$$>$0.2~GeV$/c$ and $|\eta|$$<$1 as a function of the leading jet $p_{T}$. In this and the following figures, data points are placed at the center of each $p_{T}$ bin. The statistical uncertainties are shown as vertical bars, which are smaller than the data symbols. The box heights are systematic uncertainties, while their widths correspond to the $p_{T}$ bin sizes. The Toward and Away average multiplicity densities, $\langle \frac{d N_{ch}}{d\eta d\phi}\rangle$, both show a rise with leading jet $p_{T}$. For the Transverse region, $\langle \frac{d N_{ch}}{d\eta d\phi}\rangle$ tends to slightly decrease as the leading jet $p_{T}$ increases. In contrast, at LHC energies the mid-pseudorapidity transverse multiplicity was observed to quickly increase and then saturate or slightly increase with increasing leading jet/track $p_{T}$ \cite{CMS_UE_arxiv1507.07229, ATLAS_UE_arxiv1701.05390}. PYTHIA 6 (STAR), PYTHIA 6 default Perugia 2012 and PYTHIA 8 default Monash 2013 simulations are also shown as curves with widths from widest to thinnest, respectively.  
Deviations for all the simulations from data for jet $p_{T}$$>$15~GeV$/c$ are observed in the Transverse region, with PYTHIA 6 (STAR), which was tuned with STAR published minimum bias spectra, closest to the measured results. However, while the agreement between data and  PYTHIA 6 (STAR) is reasonable, additional improvements to the  tuning or modeling itself would still be appropriate at RHIC energies.

\begin{figure}[htb]
\centering
\includegraphics[width=0.4\textwidth]{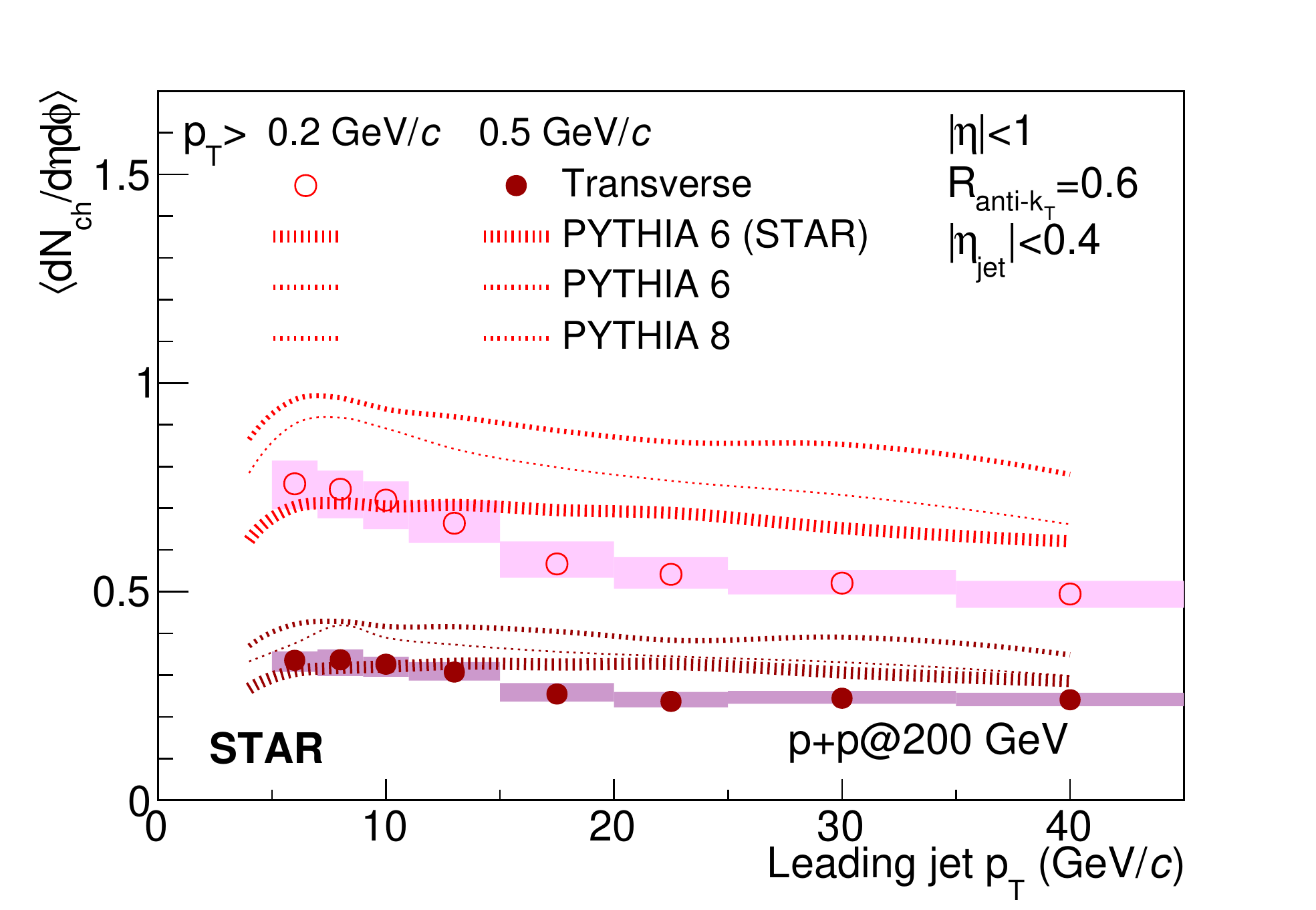}
\caption{Transverse region average charged particle densities for $p_{T}$$>$0.2~GeV/$c$ (open symbols) and $p_{T}$$>$0.5~GeV/$c$ (filled symbols). Simulations are also shown as curves.  The wide curves are PYTHIA 6 (STAR). The middle width curves are default PYTHIA 6 Perugia 2012 tune. The thin curves are PYTHIA 8 Monash 2013 tune.  
 }
\label{fig:mult2}
\end{figure}

To allow comparison with results obtained at facilities with higher collision energies, analyses for particle $p_T$$>$ 0.5~GeV/$c$ were also performed. Figure~\ref{fig:mult2} compares the fully corrected $\langle \frac{d N_{ch}}{d\eta d\phi}\rangle$ in the Transverse region as a function of the leading jet $p_{T}$ for particle $p_{T}$$>$0.2~GeV/$c$ and $p_{T}$$>$0.5~GeV/$c$.  Similar trends are observed for these two $p_{T}$ cases, with mismatch between data and PYTHIA.

\begin{figure}[htb]
\centering
\includegraphics[width=0.4\textwidth]{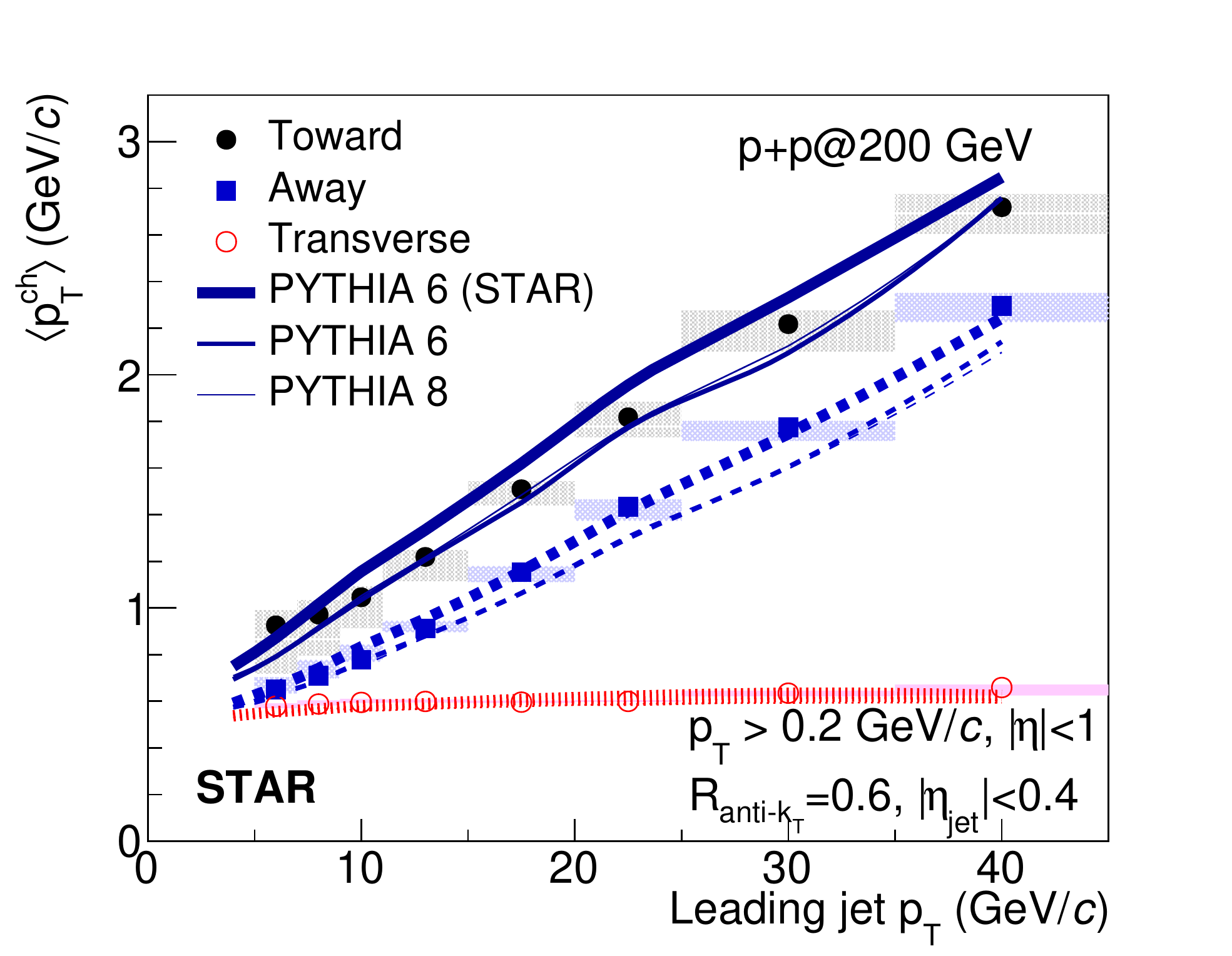}
\caption{Charged particle $\langle p_{T}\rangle$  for Toward, Away, and Transverse regions as functions of the leading jet $p_{T}$, with charged particle $p_{T}$$>$0.2~GeV/$c$. Simulations are also shown as curves. The wide curves are PYTHIA 6 (STAR). The middle width curves are default PYTHIA 6 Perugia 2012 tune. The thin curves are PYTHIA 8 Monash 2013 tune.  Note the three curves overlap for the Transeverse region calculations. }
\label{fig:pt}
\end{figure}

The $\langle p_{T}\rangle$ was measured to further profile the characteristics of the underlying event. Figure~\ref{fig:pt} shows the fully corrected charged particle $\langle p_{T}\rangle$ as a function of the leading jet $p_{T}$ for the three regions, with particle $p_{T}$$>$0.2~GeV/$c$. The Transverse region $\langle p_{T}\rangle$ slightly increases as the leading jet $p_{T}$ increases. Both the Toward and Away regions  show linearly increasing trends. PYTHIA simulations, shown as curves, provide a better description of the $\langle p_{T}\rangle$ measurements than the average multiplicity density.

\begin{figure}[htb]
\centering
\includegraphics[width=0.4\textwidth]{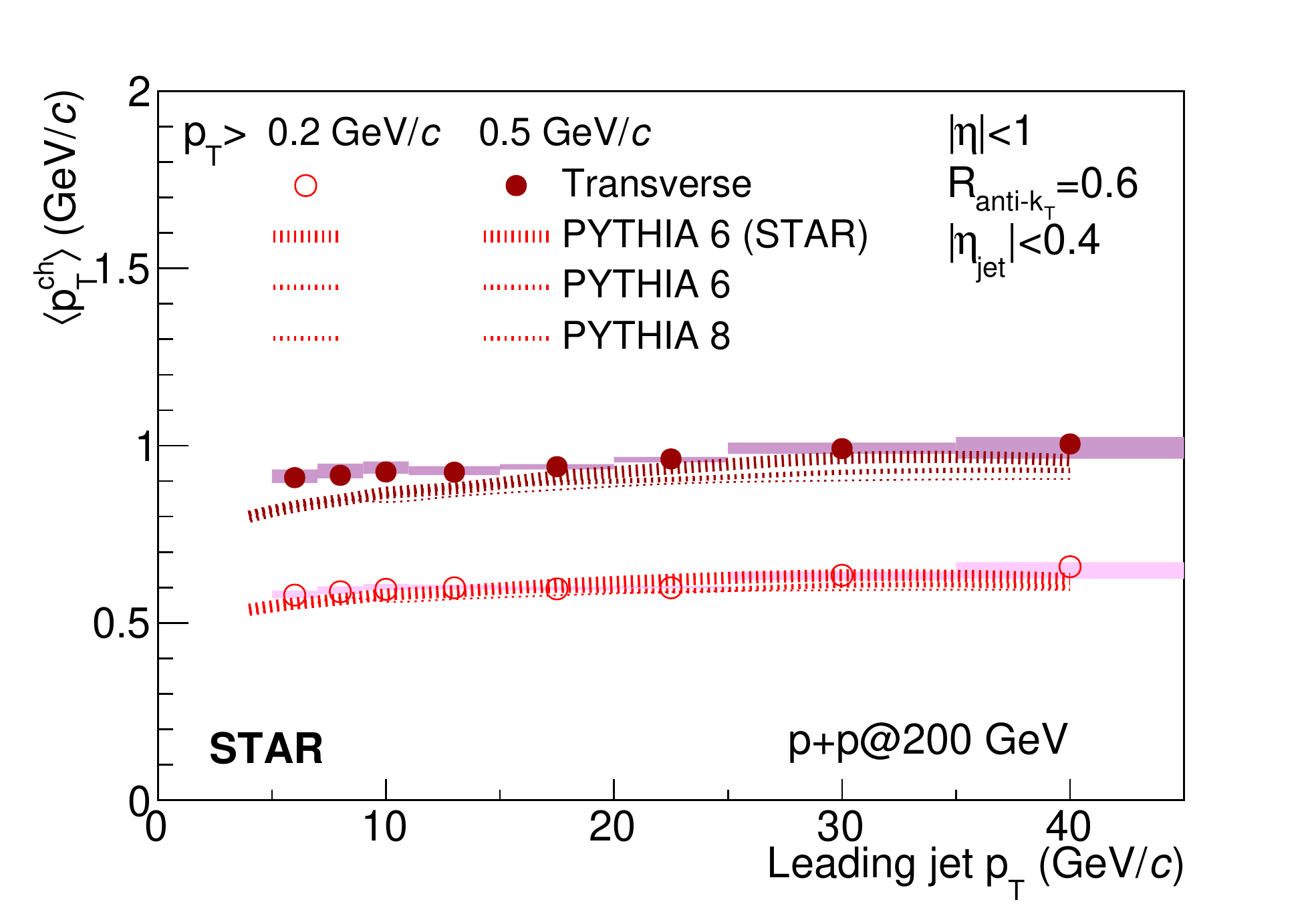}
\caption{Transverse region charged particle $\langle p_{T}\rangle$ as a function of the leading jet $p_{T}$ for $p_{T}$$>$0.2~GeV/$c$ (open symbols) and $p_{T}>0.5$~GeV/$c$ (filled symbols). Simulations are also shown as curves. The wide curves are PYTHIA 6 (STAR). The middle width curves are default PYTHIA 6 Perugia 2012 tune. The thin curves are PYTHIA 8 Monash 2013 tune.   }
\label{fig:pt2}
\end{figure}

Figure 5 shows the fully corrected transverse region charged particle $\langle p_{T}\rangle$ as a function of the leading jet $p_{T}$ for $p_{T}$$>$0.2~GeV/$c$ and $p_{T}$$>$0.5~GeV/$c$. Similar trends are observed for these two $p_{T}$ cases.

\begin{figure}[htb]
\centering
\includegraphics[width=0.4\textwidth]{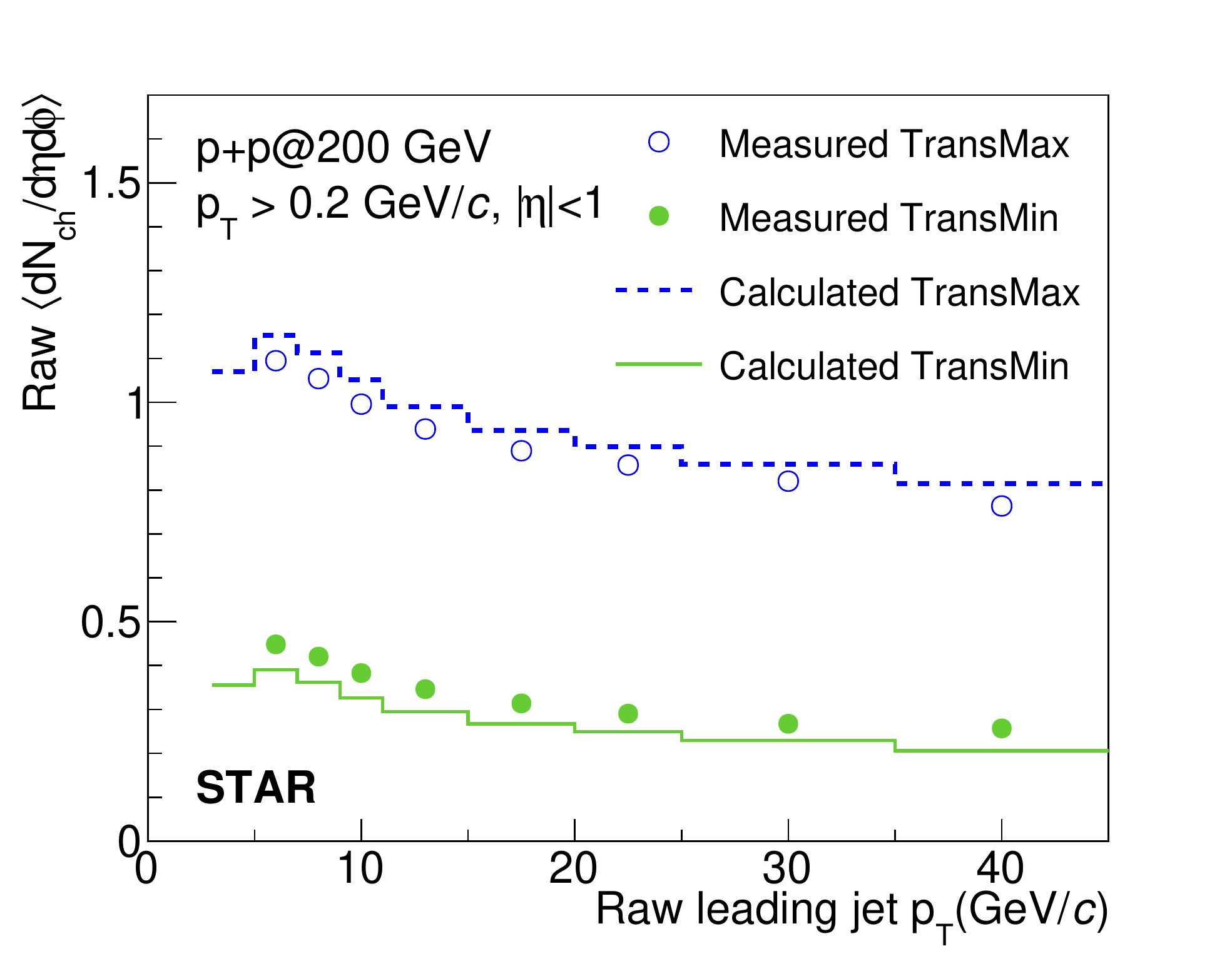}
\caption{Detector-level average charged particle multiplicity densities for TransMax and TransMin as functions of the leading jet $p_{T}$. Points are measured data. Histograms are calculated values assuming TransMax and TransMin are derived from the same parent distribution. See text for details.  }
\label{fig:TransMaxVsMin}
\end{figure}

Figure~\ref{fig:TransMaxVsMin} shows the uncorrected  detector-level TransMax and TransMin average charged particle multiplicity densities as a function of the leading jet $p_{T}$.  At $\sqrt s $ = 200~GeV, we observe that while the multiplicity densities in the TransMax and TransMin regions have different overall magnitudes, they both exhibit slightly decreasing trends with increasing leading jet p$_{T}$ with the difference in the densities staying roughly constant. Meanwhile, experiments at the LHC have reported an increasing difference in the charged particle multiplicities of these regions with increasing leading trigger $p_{T}$~\cite{CMS_UE_arxiv1507.07229, ATLAS_UE_arxiv1701.05390}. 
Increasing contributions to the TransMax region from wide angle third-jet production via ISR/FSR is understood as the main physical cause of the LHC results. To further investigate our results, we tested the hypothesis that the TransMax and TransMin charged particle density distributions are drawn, statistically independently, from the same parent probability distribution, $f(x)$. In such a case the probability distribution functions for the measured charged particle multiplicities in the TransMax and TransMin regions,  $f^{max}(x)$  and $f^{min}(x)$ respectively, can be expressed as $f^{max}(x) = 2 f(x) F(x)$  and $f^{min}(x)  = 2 f(x) (1-F(x))$, where $F(x)$ is the cumulative distribution of $f(x)$.

The calculated averages of $f^{max}(x)$ and $f^{min}(x)$ as a function of leading jet p$_{T}$ are shown as curves in Fig.~\ref{fig:TransMaxVsMin}, when $f(x)$ is taken as the average of the measured TransMax and TransMin distributions.

These calculated distributions account for most of the features of the measured data, suggesting any net contribution from additional physical sources is small. Detector-level variables were compared here to avoid the impact of additional fluctuations from the unfolding procedure, with the caveat that any detector response asymmetries could enlarge the TransMax and TransMin difference at the detector-level. 
This suggests that there are less ISR/FSR contributions to $p$+$p$ events at RHIC energies than at LHC energies.

\begin{figure}[htb]
\centering
\includegraphics[width=0.5\textwidth]{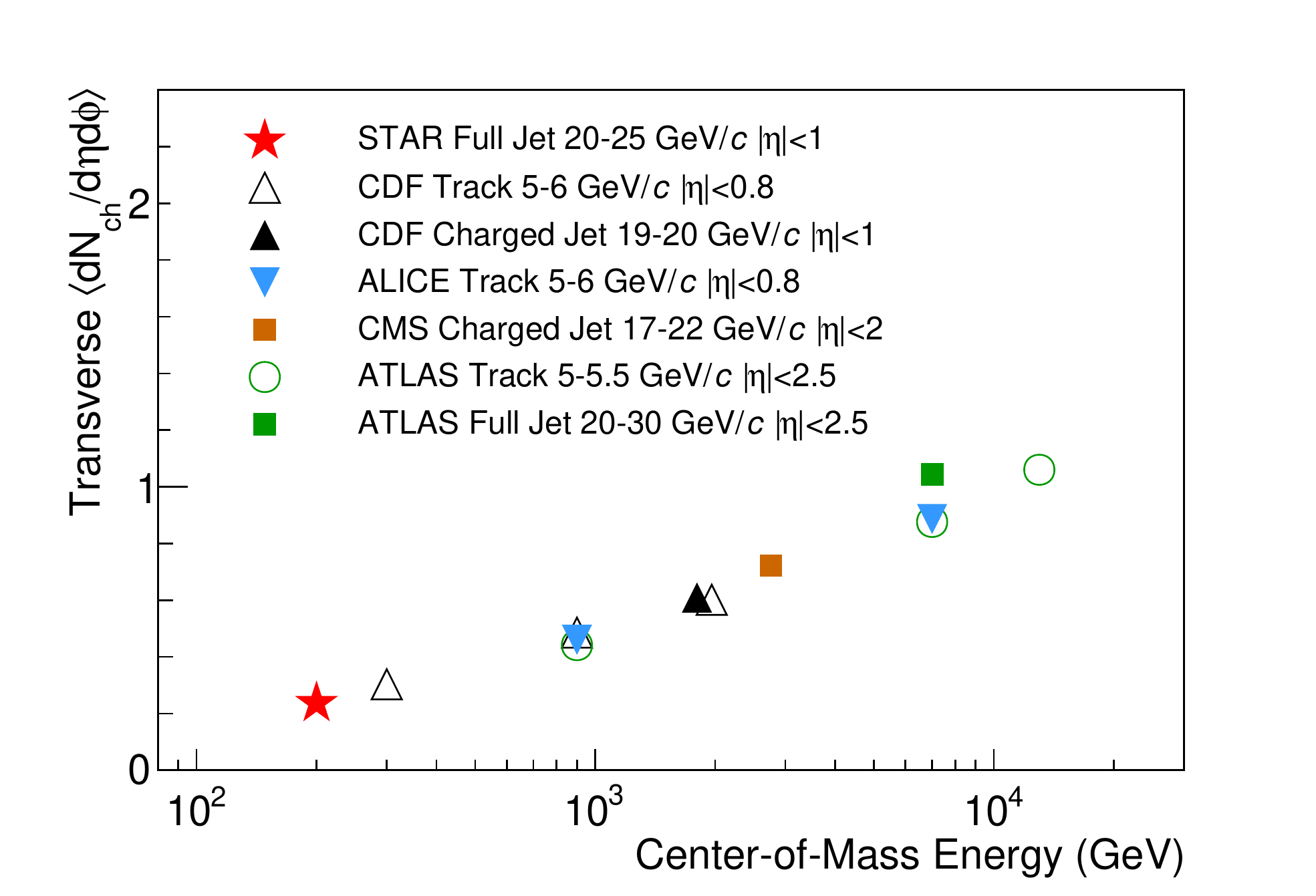}
\caption{Transverse charged particle densities at various center-of-mass collision energies. Uncertainties smaller than the marker sizes. See text for details.}
\label{fig:EneDep}
\end{figure}

Besides the underlying event activity's dependence on the hard scattering reference energy scale, studying the collision energy dependence also contributes to our understanding of how to model these low momentum processes. Detailed underlying event measurements have been reported at a variety of collision energies, which makes it possible to study the energy scaling of the underlying event phenomena, including a possible collision energy dependence of the transverse momentum cut-off between hard and soft scatterings \cite{Hera2LHC}. A world-data comparison with our RHIC measurements is done with jets or charged particles as the hard scattering reference to avoid any possible complexity that different hard references may introduce to the underlying event activity.   Since the underlying event multiplicities at the Tevatron and LHC energies show the same general trend of a rapid rise with reference p$_{T}$  before reaching a plateau~\cite{CMS_UE_arxiv1507.07229}, comparison points were chosen from the region where the plateau is just reached. An exception is the ATLAS full jet measurement at 7 TeV, where the general trend of the underlying event is not observed due to limited kinematic reach at lower $p_{T}$. In this case, the lowest $p_{T}$ data point was chosen for comparison \cite{ATLAS_UE_EPJC74_2014_2965}. Figure~\ref{fig:EneDep} shows the collision energy dependence of the charged particle density extracted for these Transverse regions and  Table ~\ref{Table:experiments} details the beam energy, trigger type and $p_{T}$ range, and  Transverse region's $\eta$ coverage for each experiment. Note all the reported results use charged particles with $p_{T}$$>$0.5~GeV/$c$ to measure the underlying event (UE) activity and that the LHC and STAR results are from $p$+$p$ collisions while those of CDF are from $p$+$\bar{p}$ events.

A near-linear increase with the log of the collision energy is observed in Fig.~\ref{fig:EneDep}, despite the different references and pseudorapidity coverage for the various measurements. It has been shown in \cite{PYTHIA6tune_arxiv1005.3457} (for example), that it remains unclear how to model correctly the collision energy scaling of the $p$+$p$ data;  a tension emerges when trying to simulate minimum bias and high multiplicity data with the same tune. This study of the collision energy dependence of the underlying event's average charged particle density provides additional information that may aid in resolving this tension. 

\begin{table}[htb]
\centering
\caption{The collaboration, beam energy, trigger types and $p_{T}$ ranges, and Transverse $\eta$ ranges used for the data plotted in Fig.~\ref{fig:EneDep}.}

\label{Table:experiments}
\begin{tabular}{M{1.5cm}  M{1.2cm} M{2cm} M{1.9cm} M{1.4cm} }
\hline
\hline
 \textbf{Collab.} & \textbf{$\sqrt{s}$ (GeV)} & \textbf{ Trigger Type}& \textbf{Trigger Range (GeV/$c$)} & \textbf{UE $|\eta|$ Range} \\ \hline
STAR & 200  & Full Jet $R_{\text anti-k_{T}}$=0.6 & 20$<$$p_{T}$$<$25 &1 \\ 
 CDF \cite{CDF_PRD92}& 300 & Charged particle & 5$<$$p_{T}$$<$6 & 0.8  \\ 
 CDF \cite{CDF_PRD92}& 900 & Charged particle & 5$<$$p_{T}$$<6$ & 0.8  \\ 
 CDF \cite{CDF_PRD92} & 1960 & Charged particle & 5$<$$p_{T}$$<$6 & 0.8 \\ 
 CDF \cite{CDF_PRD65_092002} & 1800 & Charged Jet $R_{\text cone}$=0.7& 19$<$$p_{T}$$<$20 &  1\\ 
 
 ATLAS \cite{ATLAS_UE_PRD83_2011_112001} & 900 & Charged particle & 5$<$$p_{T}$$<$5.5  & 2.5 \\ 
 ATLAS  \cite{ATLAS_UE_PRD83_2011_112001}& 7000 & Charged particle & 5$<$$p_{T}$$<$5.5 & 2.5 \\
 ATLAS  \cite{ATLAS_UE_arxiv1701.05390}& 13000 & Charged particle & 5$<$$p_{T}$$<$5.5 & 2.5  \\
 ATLAS  \cite{ATLAS_UE_EPJC74_2014_2965} & 7000 & Full Jet $R_{\text anti-k_{T}}$=0.4& 20$<$$p_{T}$$<$30 & 2.5 \\ 
 
ALICE  \cite{ALICE:2011ac} & 900 & Charged particle & 5$<$$p_{T}$$<$6 & 0.8 \\ 
 ALICE  \cite{ALICE:2011ac} & 7000 & Charged particle & 5$<$$p_{T}$$<$6 & 0.8 \\ 

 CMS \cite{CMS_UE_arxiv1507.07229} & 2760 & Charged jet  $R_{\text SISCone}$=0.5& 17$<$$p_{T}$$<$22 & 2 \\ 
 \hline
\hline

\end{tabular}

\end{table}

\section{\label{sec:sum}Summary}
We have reported several observables sensitive to the underlying event activity in $p$+$p$ collisions at $\sqrt{s}$ = 200~GeV as recorded by the STAR experiment. The results used full jets reconstructed at mid-pseudorapidity $|\eta_{jet}|$$<$0.4 with $R_{\text anti-k_{T}}$=0.6 and 5$<$$p_{T}^{jet}$$<$45~GeV/$c$. The charged particles used for the underlying event measurements were required to have $|\eta|$$<$1, 0.2$<$$p_{T}$$<$20~GeV/$c$, and $60^{\circ}$$\le$$|\Delta\phi|$$\le$$120^{\circ}$. The reported observables were corrected to the generator-level to enable direct comparison with theoretical calculations. The detector response corrections were performed via a data-driven trigger correction and embedded PYTHIA 6 (STAR)+GEANT simulations into zero-bias experimental data. The uncertainties from the detector response corrections are included in the systematic uncertainties. 

The corrected observables were reported for three topological regions, Toward, Away, and Transverse, based on the azimuthal angle of the particles relative to that of the highest $p_T$ jet. The average charged particle multiplicity densities and their mean transverse momenta in these regions were studied as functions of the leading jet $p_{T}$. These correlations characterize the relationship between the underlying event activity and a hard scattering in $p$+$p$ collisions. The Transverse charged particle density was observed to slightly decrease with the leading jet $p_{T}$, while the Transverse $\langle$$p_{T}$$\rangle$ slightly increases with the leading jet $p_{T}$. The slight negative correlation  of  $ \langle$$\frac{dN_{ch}}{d\eta d\phi}$$\rangle$ is consistent with energy conservation restricting particle production in the Transverse region as the leading jet becomes more energetic.

The Transverse regions were further split into TransMax and TransMin areas. The observation of similar charged particle production in the TransMax and TransMin regions at RHIC, which is not observed at the LHC, suggests that ISR/FSR contributes less to the underlying event at $\sqrt{s}$ = 200~GeV than at TeV collision energies. The results reported here are therefore predominantly sensitive to soft MPI and beam-remnant activity. 

An energy dependence study of the Transverse region charged particle density shows a near-linear increase with the log of the collision energy, which contributes to the understanding of the energy scaling of the underlying event dynamics. These underlying event activity measurements, in combination with the previously reported minimum bias observables \cite{STAR_PLB616_8_16, STAR_PRL108_072302}, provide valuable input and constraints to the predominantly phenomenological modeling in Monte Carlo event generators of the low-momentum QCD processes in $p$+$p$ collisions.

We thank the RHIC Operations Group and RCF at BNL, the NERSC Center at LBNL, and the Open Science Grid consortium for providing resources and support.  This work was supported in part by the Office of Nuclear Physics within the U.S. DOE Office of Science, the U.S. National Science Foundation, the Ministry of Education and Science of the Russian Federation, National Natural Science Foundation of China, Chinese Academy of Science, the Ministry of Science and Technology of China and the Chinese Ministry of Education, the National Research Foundation of Korea, Czech Science Foundation and Ministry of Education, Youth and Sports of the Czech Republic, Hungarian National Research, Development and Innovation Office, New National Excellency Programme of the Hungarian Ministry of Human Capacities, Department of Atomic Energy and Department of Science and Technology of the Government of India, the National Science Centre of Poland, the Ministry  of Science, Education and Sports of the Republic of Croatia, RosAtom of Russia and German Bundesministerium fur Bildung, Wissenschaft, Forschung and Technologie (BMBF) and the Helmholtz Association.

\bibliography{ppUE200} % Tell bibtex which .bib file to use (this one is some example file in TexLive's file tree)

\end{document}